\documentclass[conference]{IEEEtran}
\usepackage{url}
\usepackage{cite}
\usepackage{amsmath,amssymb,amsfonts}
\usepackage{algorithmic}
\usepackage{graphicx}
\usepackage{textcomp}
\usepackage{xcolor}
\usepackage{float}
\usepackage{svg}
\def\BibTeX{{\rm B\kern-.05em{\sc i\kern-.025em b}\kern-.08em
    T\kern-.1667em\lower.7ex\hbox{E}\kern-.125emX}}
\begin{document}

\title{Benchmarking Energy Efficiency of Large Language Models Using vLLM\\
\thanks{Corresponding author: kallepronk@proton.me}
}

\author{\IEEEauthorblockN{1\textsuperscript{st} Kalle Pronk}
\IEEEauthorblockA{\textit{Allumni Master Applied IT} \\
\textit{Fontys University of Applied Sciences}\\
Eindhoven, The Netherlands \\
kallepronk@proton.me}
\and
\IEEEauthorblockN{2\textsuperscript{nd} Qin Zhao}
\IEEEauthorblockA{\textit{Sustainable Data \& AI Group} \\
\textit{Fontys University of Applied Sciences}\\
Eindhoven, The Netherlands \\
q.zhao@fontys.nl}
}

\maketitle

\begin{abstract}
The prevalence of Large Language Models (LLMs) is having an growing impact on the climate due to the substantial energy required for their deployment and use. To create awareness for developers who are implementing LLMs in their products, there is a strong need to collect more information about the energy efficiency of LLMs. While existing research has evaluated the energy efficiency of various models, these benchmarks often fall short of representing realistic production scenarios. In this paper, we introduce the LLM Efficiency Benchmark, designed to simulate real-world usage conditions. Our benchmark utilizes vLLM, a high-throughput, production-ready LLM serving backend that optimizes model performance and efficiency. We examine how factors such as model size, architecture, and concurrent request volume affect inference energy efficiency. Our findings demonstrate that it is possible to create energy efficiency benchmarks that better reflect practical deployment conditions, providing valuable insights for developers aiming to build more sustainable AI systems.
\end{abstract}


\section{Introduction}
\label{sec:Intro}

Large Language Models (LLMs) have seen a significant rise in popularity in recent years. They are increasingly integrated into everyday applications, such as Google’s AI-generated summaries for search results, OpenAI’s GPT-4o, and the growing adoption of AI agents across various platforms. The industry’s pursuit of more accurate and capable models has led to a dramatic increase in the size of LLMs. This trend is evident in the evolution of OpenAI’s GPT series. GPT-1 contained 117 million parameters, while GPT-3, released just three years later, scaled up to 175 billion parameters. GPT-4 is estimated to exceed one trillion parameters, highlighting the exponential growth in model complexity.

As models grow in size, their energy consumption increases accordingly. The widespread adoption of LLMs has led to a significant rise in energy usage and associated CO$_2$ emissions, raising concerns about their environmental impact.

Energy consumption and carbon emission both from the training of LLMs, as well as from the inference have gained attention in academia \cite{Argerich_Patiño-Martínez_2024, 10.1145/3531146.3533234}. However, most recent research in inference has been limited in collecting energy consumption in the lab conditions \cite{Argerich_Patiño-Martínez_2024, 10.5555/3455716.3455964}. By simulating models in settings that do not adapt up-to-date tooling, that do not reflect how modern LLM services serve text generation and other application.

We introduce the LLM Efficiency Benchmark, designed to evaluate LLMs under realistic conditions that reflect modern serving environments. This benchmark leverages vLLM \cite{vLLM}, a high-performance LLM backend optimized for handling large volumes of requests. vLLM improves model throughput by optimizing memory management and GPU utilization, making it well-suited for simulating production-level workloads.

The remainder of this paper is organized as follows: Section \ref{sec:Related Work} reviews related work. Section \ref{sec:Methodology} outlines the design of the benchmark along with the hardware and software configurations used in our experiments. In Section \ref{sec:Results}, we present our results, analyzing how LLM energy efficiency varies across different scenarios. Section \ref{sec:Discussion} discusses how our findings relate to prior research and offers recommendations for future efforts in developing realistic benchmarks. Finally, Section \ref{sec:Conclusion} concludes the paper. 

\section{Related work}
\label{sec:Related Work}

AI models have taken a exponential leap in complexity ever since Alex-Net came out in 2012 \cite{openai2024}. The increased complexity of AI models are often a product of trying to achieve more accuracy. Less attention has been directed at the carbon offset of these models. This section will discuss how carbon offset estimations can be made based on energy consumption, how computer energy consumption can be measured, what earlier researches have measured Large Language Model power consumption, and how vLLM serves as an efficient LLM serving backend.

\subsection{Carbon Offset}
An AI model's carbon offset can be estimated by measuring its power consumption and multiplying it with its local power grid's carbon intensity. This can be location and time dependent, as power grids can have different power sources such as solar, hydro, wind, coal or gas power stations. Another factor necessary to estimate the carbon offset of AI models is its Power Usage Effectiveness (PUE). This coefficient accounts for the extra infrastructure power draw from the data center or other location the model is being run on \cite{Strubell_Ganesh_McCallum_2020}. Carbon offset can best be represented with CO$_2eq$. This measures the effect of green house gasses by the equivalent amount of CO$_2$ emissions needed to achieve the same effect. For instance, methane gas has 25 times more global warming potential then CO$_2$, thereby making it 25 CO$_2eq$\cite{bloom}.

Model carbon offset can be reduced without having to reduce the power consumption of said model. One way to achieve this is by changing the location of the data center. One study finds that running a model in Quebec would result in 30 times less carbon emissions than running the same model in Estonia \cite{10.5555/3455716.3455964}.
Another way lower carbon emissions can be achieved without lowering power consumption is by optimizing the hours during the day the model is active. Due to weather dependent energy sources like solar panels and windmills, power grids can have varying carbon intensities throughout the day. The Pause and Resume algorithm can reduce energy consumption in ideal conditions by nearly 30\% \cite{10.1145/3531146.3533234}.

These strategies are useful in the model training phase, but less so during the inference phase. Data privacy regulations, cost and latency issues become problems when choosing server locations for serving model inference. Furthermore, modern LLM serving solutions require the service to always be available, unlike training that can be done in batches.

\subsection{Energy consumption measurement}
Estimating carbon offset requires accurate energy measurements. When there is physical access to the machine this can be done quite easily using a power meter or power monitor. However, tracking the energy consumption of a single process is more difficult \cite{Argerich_Patiño-Martínez_2024}.
Most embedded hardware comes with sensors made to track energy consumption per component, or even part of a component. These low level sensors can be accessed by tools such as NVIDIA-smi \cite{nvidiaNVIDIASystem} and powermetrics \cite{powermetrics}.

Multiple tools have been created to measure the total system energy consumption of software. All these tools use a combination of different low level interfaces like RAPL to get an accurate estimation of power consumption \cite{CodeCarbon, Eco2AI}. Most of this tooling combines power measurement with carbon estimation. This is done by tracking power grid carbon intensity worldwide. While this can give some meaningful information about the consequences of different server locations, these estimations can vary in terms of accuracy based on the provided local transparency.

\subsection{LLM Efficiency measurements}
As mentioned above, most existing research focuses on measuring and mitigating carbon offset of LLM training. The Machine Learning Emissions Calculator lets researchers measure the total carbon emissions of model training. This method relies on researchers to self report training emissions \cite{lacoste2019quantifyingcarbonemissionsmachine}, which most models on Huggingface do not do. One study finds that less than 3\% of LLM's on Huggingface report emission readings \cite{10304801}.
In many studies inference energy efficiency is estimated based on parameter size. However existing research shows no correlation between parameter amount and energy efficiency \cite{10.5555/3455716.3455964}. This is not the case when models are from the same architecture, they conclude that there is a sub-linear correlation between the two. This suggests that estimating model efficiency is possible, as long as the model architecture is the same \cite{10.5555/3455716.3455964, Argerich_Patiño-Martínez_2024}.
These studies base their metrics on experiments in lab conditions, where some pre-processing steps such as encoding are done beforehand instead of during the energy measurement. The studies also display efficiency in the energy-per-token metric. This metric is arguable as the size of a token can vary per tokenizer, and most model architectures use their own tokenizer. For instance the Gemma tokenizer \cite{lunaryGemmaTokenizer} encodes text on a per word basis, while the GPT-3 tokenizer \cite{GPTtokenizer} can process a word into multiple tokens. The existing studies implement the experiments using scripts made with PyTorch and Huggingface's Transformers library \cite{10.5555/3455716.3455964, Argerich_Patiño-Martínez_2024}. This can provide tests with fine-grain control of the variables, but misses out on modern serving technology that greatly optimizes model hardware performance.

\subsection{LLM serving}
In order to provide results that simulate real life situations, LLM serving techniques need to be taken into account. Scaling real-time LLM inference serving can be a resource intensive task. Modern back-ends can optimize model serving to be production ready. This can be done through a number of features, such as hardware acceleration, automatic batching, and memory optimization.
One of these back-ends is vLLM \cite{vLLM}. vLLM serves models with high throughput and low latency through efficient use of hardware resources. It achieves this with its paged-attention algorithm which allows values to be stored in non continuous sequences, thereby making memory management more flexible.

\section{Methodology}
\label{sec:Methodology}

The methodology covers the design of the LLM Efficiency Benchmark and the design of the experiments. The section is divided into the following subsections: the benchmark's energy consumption measurement strategy, the benchmark's design, and the experiments test parameters. Each part is described in the following.

\subsection{Measurement strategy}
Measuring power consumption of hardware components was done through CodeCarbon \cite{CodeCarbon}. CodeCarbon estimates power consumption per computer component by reading low-level sensors through programs like NVIDIA-smi and Intel Powergadget every 15 seconds. By multiplying the voltage measurements with the time interval, CodeCarbon can estimate power consumption on a per-component basis.

\subsection{Benchmark design}
This subsection describes the design of the software used to execute a set of tests in order to run the benchmark. vLLM has been chosen for running inference because of its high performance and open source nature \cite{vLLM}.

The test is made up of multiple runs. Each run has a different set of parameters, these are: model, request amount and request rate. Before each run, 200 warm up requests are sent to the backend. This will prevent variation in the measurements due to the hardware not reaching heat saturation. All tests were done with no request rate. This means that all requests are sent to the backend at the same time.
The specific requests to be sent are selected from a dataset. The dataset can be selected per run. Every run will go through the dataset in the same order.
Before the requests were sent, the CodeCarbon tracker was activated. The tracker stayed active until all requests were returned by the backend.

\subsection{Test parameters}
This subsection describes the parameters and scenarios used for the evaluation of LLM power consumption during inference. All tests are done on a PC with 2 NVIDIA GeForce RTX 3090's a 13th Gen Intel(R) Core(TM) i9-13900K processor and 128 Gigabytes of DDR5 RAM.

All tests are preformed with phrases from the HellaSwag dataset \cite{zellers2019hellaswagmachinereallyfinish}. HellaSwag is a dataset designed to test natural language accuracy. It presents the model with uncompleted phrases, and asks the model to complete these sentences. The model's results will not be collected, as accuracy is not in the focus of this experiment.

There will be an analysis of how LLMs preform under these different parameters:

\subsubsection{Request load}
As vLLM is designed to handle requests at high volume, it is imperative to measure wether the efficiency changes when there are different request amounts. By preforming all further experiments in request loads of 5 to 5000 at a time, it is possible to analyze how model efficiency changes throughout these different loads.

\subsubsection{Model size:}
The size of a model is most accurately measured by the number of trainable or free parameters. This directly affects the amount of operations that have to be performed during inference as well as the amount of memory needed to load the model. 

To analyze the differences in energy consumption across model sizes, we chose the Pythia family of models \cite{biderman2023pythiasuiteanalyzinglarge}. The list of selected models can be found in Table \ref{fig:model size params}. Pythia has been chosen not to optimize each model's size to its greatest potential, instead, the family of models has been designed to isolate the model's parameter amount as a variable. This provides a solid basis to perform size based analysis.

\begin{table}[H]
\centering
\caption{List of the selected Pythia models. Layers are sequential steps in which parameters are calculated. One layer cannot be started until the previous has been completed.}

    \begin{tabular}{|l|l|l|l|l|}
        \hline
        \multicolumn{1}{|c|}{\textbf{Params}} & \multicolumn{1}{c|}{\textbf{layers}} \\ \hline
        70M & 6  \\ \hline
        160M & 12  \\ \hline
        410M & 24  \\ \hline
        1B & 16  \\ \hline
        1.4B & 24  \\ \hline
        2.8B & 32  \\ \hline
        6.9B & 32  \\ \hline
        \end{tabular}
\label{fig:model size params}

\end{table}
\subsubsection{Model Architecture:}
The architecture of a model refers to the implementation strategy of components like embedding dimensions, number of layers or the number of attention heads. Most "base" models often have their own architecture. With different quantizations or parameter sized versions of that architecture being considered to be in the same "family".
To find suitable models to compare architectures, they need to be the same parameter size. The list of selected models can be found in Table \ref{fig:model architecture}. The models have been selected to allow for cross-comparison with an other study \cite{Argerich_Patiño-Martínez_2024}.

\begin{table}[H]
\centering
\caption{List of the selected models for testing architecture.}

    \begin{tabular}{|l|l|}
        \hline
        \multicolumn{1}{|c|}{\textbf{Name}} & \multicolumn{1}{|c|}{\textbf{Params}}  \\ \hline
        Pythia \cite{biderman2023pythiasuiteanalyzinglarge} & 2.8B  \\ \hline
        Dolly V2 \cite{databricksFreeDolly} & 2.8B  \\ \hline
        BLOOM \cite{workshop2023bloom176bparameteropenaccessmultilingual} & 3B \\ \hline
        Redpajama \cite{togetherReleasingRedPajamaINCITE}& 2.8B  \\ \hline
        \end{tabular}
\label{fig:model architecture}
\end{table}

\section{Results}
\label{sec:Results}
The results address three factors influencing energy consumption: the number of requests, the size of the model and the model's architecture.

\subsection{Request amount}
\begin{figure}[H]
    \centering
    \includegraphics[width=1\linewidth]{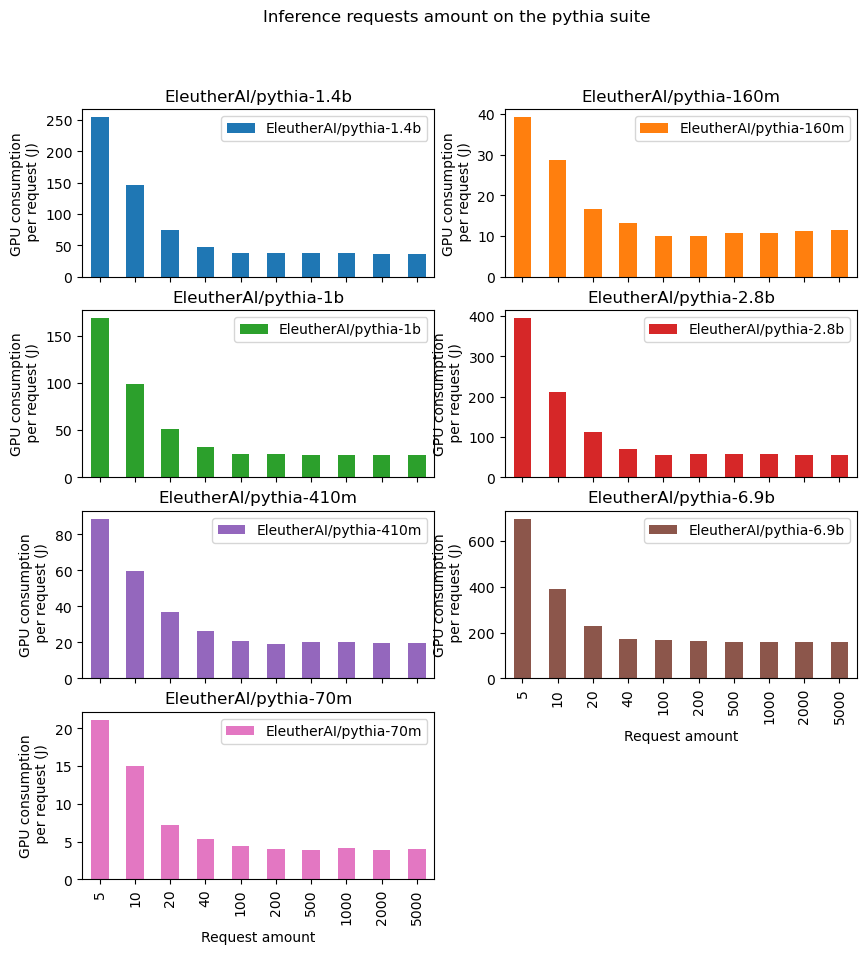}
    \caption{Inference energy cost per request on the Pythia suite. model size varies from 70 million to 6.9 billion parameters.}
    \label{fig:request-pythia}
\end{figure}

The amount of requests being sent to the serving backend at the same time has effect on the energy consumption per request as shown in Figure \ref{fig:request-pythia}. This Figure presents seven graphs, each graph shows the results of one of the selected models from Table \ref{fig:model size params}. The X-axis represents the request amount. This is the amount of requests that was send to vLLM at the same time. The Y-axis represents the GPU consumption per request as measured in joules. As the request amount increases, the consumption per request decreases. This behavior stops at 100 simultaneous requests. Beyond this measurement, the per request energy costs reach a plateau. Larger models have less variance in energy consumption per request amount. For instance, the 6.9 billion parameter version of Pythia shows a stabilized energy consumption per request at 40 parameters.

\begin{figure}[H]
    \centering
    \includegraphics[width=1\linewidth]{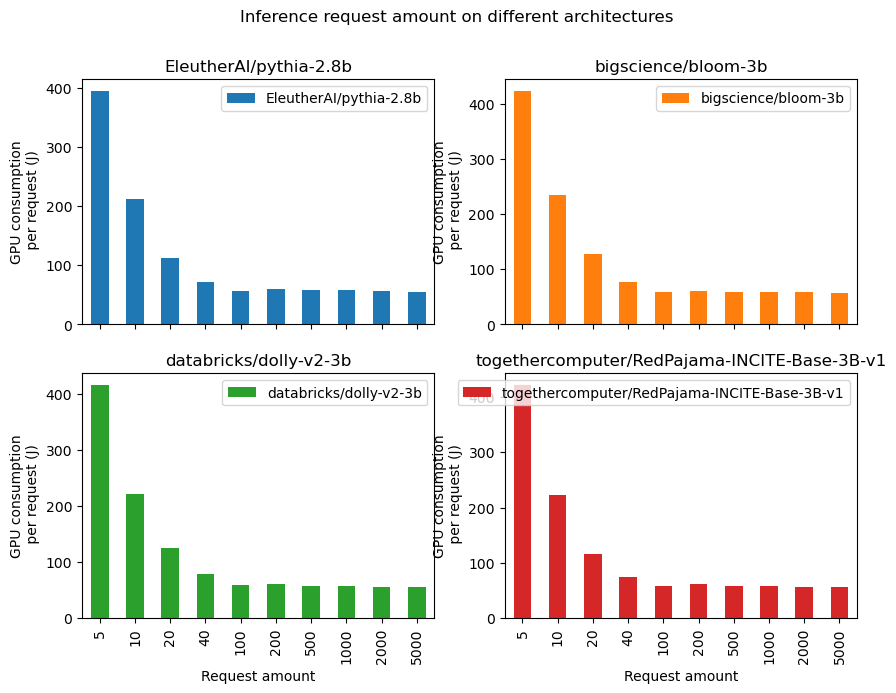}
    \caption{Inference energy cost per request on different model architectures. Each model is close to 3B parameters in size.}
    \label{fig:request-3B}
\end{figure}

The same behavior is true for models of different architectures as shown in Figure \ref{fig:request-3B}. This figure presents four experimental results of the models listed in Table \ref{fig:model architecture}, the X- and Y-axis are the same as in Figure \ref{fig:request-pythia}. As before, it shows that energy consumption per request decreases until it reaches a plateau at 100 simultaneous requests.

\subsection{Model size}
\begin{figure}[H]
    \centering
    \includegraphics[width=1\linewidth]{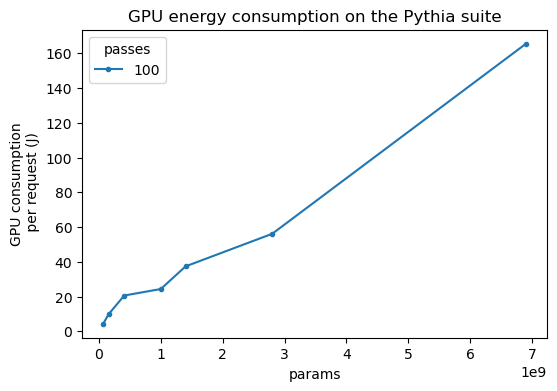}
    \caption{Inference energy costs per 100 requests on the Pythia suite. There is a close to linear relationship between parameter size and GPU energy consumption per request.}
    \label{fig:linear}
\end{figure}

The energy per request is observed to increase as model parameter size increases as shown in Figure \ref{fig:linear}. This graph presents the energy consumption of a model in joules and is calculated by dividing the total energy consumption of the GPU during the experiment by the amount of requests processed. 100 requests are used, as they have been shown to read plateau in Figure \ref{fig:request-pythia} and Figure \ref{fig:request-3B}.

The close to linear relationship is shown to be true for most models except the 410 million parameter model which consumes roughly the same amount of energy per request as the 1B parameter model.

\subsection{Model Architecture}
\begin{figure}[!h]
    \centering
    \includegraphics[width=1\linewidth]{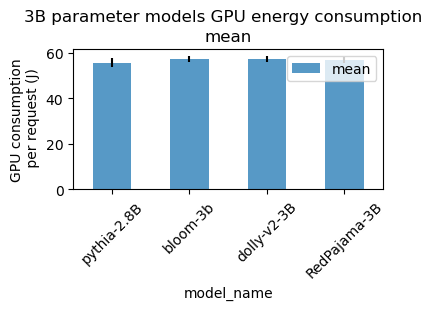}
    \caption{Inference energy costs per 100 requests on different 3B parameter models. Each model has been tested 10 times, the black line shows the standard deviation of the tests.}
    \label{fig:3B}
\end{figure}

There are no significant differences between models of comparable size but different architecture. As depicted in Figure \ref{fig:3B}, four models with size close to 3B parameters are chosen. Their GPU consumption per request are compared, and are with marginal differences.

\section{Discussion}
\label{sec:Discussion}

In our benchmarks with vLLM we find that model efficiency decreases close to linearly with model parameter size when models are of the same architecture.  These results are in line with an earlier study from the Polytechnic University of Madrid. They found the same differences between models when comparing model parameter size to energy per token \cite{Argerich_Patiño-Martínez_2024}. However, both our study as their study found a deviation in the 1 billion and 410 million model. These models appear to have roughly the same energy efficiency, even though the parameter count has more than doubled. as stated in Argerich et al. \cite{Argerich_Patiño-Martínez_2024}, this can be explained by the difference in layers. The 1 billion parameter model has 16 layers while the 410 million model has 24 layers \cite{biderman2023pythiasuiteanalyzinglarge}. This can explain the similar power efficiency as each layer needs to be completed before another layer can start processing, thus limiting the amount that the model can be efficiently parallelized.

We also find that model architecture with similar size has no significant effect on model efficiency. This is where our results differ from Argerich et al \cite{Argerich_Patiño-Martínez_2024}. They find an significant efficiency increase of 47\% between the least and most efficient models in the category. When comparing these results, it needs to be considered that our experiments differ in multiple ways. We measure energy efficiency in joules per sent request and we use vLLM,  where they measure energy efficiency in joules per token and use the Huggingface transformers library \cite{wolf2020huggingfacestransformersstateoftheartnatural, Argerich_Patiño-Martínez_2024, 10.5555/3455716.3455964}. vLLM offers multiple optimizations that make model serving more efficient. It could be the case that vLLM's design renders the efficiency gain between architecture negligible.
While this could give a possible explanation for the different behavior, there has yet to be a clear explanation for why vLLM would offer this stabilization between architectures. Further research is needed to explore this aspect.

\subsection{Validity}
As discussed in the previous section, this research has deviated from earlier work in two main ways. The first deviation is that experiments are done through the use of vLLM for inference testing. 
vLLM is more challenging to test and validate. This is because the backend optimizes the model's inference performance in real time, by automatic batching, speculative decoding and GPU acceleration \cite{vLLM}. It creates more variables that can influence the outcome of the test. This has impact on the internal validity of the results, as the question can be posed if the tests measure the efficiency of models, or the efficiency of vLLM. It is however of great importance to use modern inference backend when testing performance as this work has tried to best replicate realistic scenario's that can be found in modern LLM serving technologies. We argue that it not possible to accurately assess a model's efficiency without also considering the serving method. Since vLLM is among the most efficient serving solutions currently available, it has been chosen for this study.
The other deviation of this work compared to previous is the use of the energy consumption per request as the metric, this is different from the energy consumption per token done by most other LLM energy efficiency papers. The preference for token based measurements is that responses and questions may be of different token sizes, thereby making tokens a consistent and accurate indicator for amount of work done. However, there are multiple arguments to be made for the energy per request metric:
\begin{enumerate}
    \item Different model architectures have different tokenizers, GPT based model might create multiple tokens per word, while other tokenizers can only encode text per individual word.
    \item Part of the LLM's structure is its tendency to create longer or shorter replies to text inputs. While this can be influenced by the models implementation, one model can create more concise answers. This can lead to efficiency gains as well.
\end{enumerate}

\subsection{Limitations}

This study has not accounted for all possible parameters, with accuracy being the most prominent omission. There are several different ways that accuracy and other metrics like perplexity can be measured. There are popular global leader-boards for LLM accuracy \cite{open-llm-leaderboard-v2}. However, these leader-boards often mainly test natural language quality or performance in specific domains. Depending on a developers use case, different accuracy metrics may hold varying degrees of relevance. For instance, a model made specially to analyze classic novels might struggle with a math accuracy test, but the math accuracy test would not be relevant in this situation. Since our focus is on energy efficiency of LLMs, we have ignored this for now.

\subsection{Future work}
\label{sec:Future Work}
This work has shown that models of comparable parameter counts are on par in terms of efficiency. Our observations differ from the earlier study when testing the impact of model architecture on energy efficiency. To reconcile these studies, further research is needed. To fill in the remaining gaps in knowledge and validity, and create more knowledge about energy efficiency of large language models, we recommend the following steps for future research:
\begin{itemize}
    \item \textbf{Test more LLM's:} To further validate results of the model parameter size and architecture, it is advised to create larger tests with more models, these can then be analyzed on parameter count and architecture/
    \item \textbf{Test more LLM serving techniques:} LLM serving tooling and techniques are evolving rapidly. It is important to test with a larger variety of vLLM competitors. Future research should look into TGI \cite{githubGitHubHuggingfacetextgenerationinference} and TensorRT \cite{githubGitHubNVIDIATensorRT} among others.
    \item \textbf{Create benchmarks on both energy efficiency and accuracy:} While this study finds a direct correlation between parameter size and energy efficiency, this does not necessarily mean that there is a correlation between energy efficiency and accuracy. As some optimized models with lower parameter counts can outperform larger models in specific accuracy benchmarks \cite{open-llm-leaderboard-v2}. Further research into energy efficiency against accuracy scores will be of great value to developers in their choice for LLM.
\end{itemize}

\section{Conclusion}
\label{sec:Conclusion}

As LLM's become increasingly ubiquitous in in our daily lives, reducing the impact they have on the climate becomes more and more relevant every day. Before developers can contribute effectively to this cause, they need a clear understanding, and be properly informed of how their choice of model influences energy consumption and overall system efficiency. Through the creation of the LLM Efficiency Benchmark, we have shown that it is possible to create a benchmark that can simulate realistic LLM serving scenario's. In the future, We will work on creating a extensive LLM energy efficiency database to support developers to create and select more energy conscious LLM's for production use.

\bibliographystyle{unsrt}
\bibliography{conference_101719}

\end{document}